\documentclass[amsmath,amssymb,amsfonts,aip,jcp,superscriptaddress,showpacs,showkeys,longbibliography,10pt,twocolumn]{revtex4-2}

\usepackage{graphicx}
\usepackage{bm}
\usepackage[english]{babel}
\usepackage{color}
\usepackage{lipsum}
\usepackage{hyperref}
\usepackage[mathlines]{lineno}
\usepackage{caption} 
\usepackage{subcaption} 

\begin{document}

\title{  
Interplay of Coil-Globule Transitions and Aggregation in Homopolymer Aqueous Solutions: Simulation and Topological Insights}

\author{Junichi Komatsu}
\affiliation{Department of Chemistry, Faculty of Science, Okayama University, 3-1-1 Tsushima-Naka, Kita-ku, Okayama 700-8530, Japan}
\author{Kenichiro Koga}\thanks{Authors to whom correspondence should be addressed: koga@okayama-u.ac.jp and jonas.berx@nbi.ku.dk}
\affiliation{Department of Chemistry, Faculty of Science, Okayama University, 3-1-1 Tsushima-Naka, Kita-ku, Okayama 700-8530, Japan}
\affiliation{Research Institute for Interdisciplinary Science, Okayama University, 3-1-1 Tsushima-Naka, Kita-ku, Okayama 700-8530, Japan}

\author{Jonas Berx}\thanks{Authors to whom correspondence should be addressed: koga@okayama-u.ac.jp and jonas.berx@nbi.ku.dk}
\affiliation{Niels Bohr International Academy, Niels Bohr Institute, University of Copenhagen, Blegdamsvej 17, 2100 Copenhagen, Denmark}

\graphicspath{{./Figures/}}

\date{\today}

\begin{abstract}
   We investigate the structural and topological properties of hydrophobic homopolymer chains in aqueous solutions using molecular dynamics simulations and circuit topology (CT) analysis. By combining geometric observables, such as radius of gyration and degree of aggregation, with CT data, we capture the relationship between coil-globule and aggregation transitions, resolving the system’s structural changes with temperature.
   Our results reveal a temperature-driven collective transition from isolated coiled chains to globular aggregates. 
    At a characteristic transition temperature $ T_c $, each chain in multichain systems undergoes a rapid coil-globule collapse, coinciding with aggregation, in contrast to the gradual collapse observed in single-chain systems at infinite dilution.
   This collective transition is reflected in geometric descriptors and a reorganization of CT motifs, shifting from intrachain-dominated motifs at low temperatures to a diverse ensemble of multichain motifs at higher temperatures. CT motif enumeration provides contact statistics while offering a topologically detailed view of polymer organization.
   These findings highlight CT’s utility as a structural descriptor for polymer systems and suggest applications to biopolymer aggregation and folding.
\end{abstract}

\maketitle
The coil-globule transition in homopolymer solutions is a conformational change from an extended coil to a compact globular state, driven \textcolor{black}{in many cases by varying temperature.\cite{Slagowski1976,Nierlich1978,Sun1980,Kubota1990,Wang1998} 
It may also be induced by changing cosolvent concentration~\cite{Zhang2001, Hao2010} or adding specific salts.~\cite{Zhang2005, Myers2024}} While this transition is ordinarily gradual in dilute systems, its {\color{black} sharpness} can vary depending on the polymer type and solution conditions.~\cite{Sun1980,Kubota1990} 

In an organic solvent, linear flexible homopolymer chains adopt a coil state at high temperatures and a globule state at low temperatures around the upper critical solution temperature.~\cite{Slagowski1976, Nierlich1978,Sun1980} The primary driving force for polymer chain collapse is typically van der Waals intrachain interactions, which are weak but become more significant at lower temperatures. 

In contrast, in aqueous solutions, water-soluble polymer chains are extended at low temperatures and globular at high temperatures.~\cite{Kubota1990, Wang1998, Maeda2000}  
The opposite temperature dependence indicates that the driving force is the hydrophobic interaction between non-polar moieties, an effective force that is weakly attractive, or could be even repulsive, at low temperatures,  but is strongly attractive at elevated temperatures. For example, the osmotic second virial coefficient of methane in water is a decreasing function of temperature, changing from a positive value near 273 K to large negative values at higher temperatures.~\cite{koga2013jpcb}
The water-mediated hydrophobic interactions are largely entropic in origin: the configurational entropy of water is greater when nonpolar solutes (moieties) are in contact with each other than when they are apart.~\cite{Widom2003} 
Such interactions play a key role in determining the native structures of biological macromolecules and in driving self-assembly into ordered forms like micelles, vesicles, or lamellar phases.~\cite{Tanford1978}

Significant effort has been devoted to studying the nature of the coil-to-globule transition of a single polymer chain, including the fully collapsed globule state, using highly dilute solutions with narrow molecular weight distributions.~\cite{Sun1980,Kubota1990,Wang1998,Maeda2000} 
At higher polymer concentrations, chain aggregation occurs at a specific temperature, making it difficult to isolate information about the single-chain coil-to-globule transition from experiments. {\color{black} Previous simulations and finite-size scaling studies of semidilute polymer solutions show that single-chain collapse and multi-chain aggregation occur at the same temperature for infinitely long chains and that
the upper consolute point for finite-length chains belongs to the three-dimensional Ising universality class~\cite{Frauenkron1997,Binder2005}.}
In contrast, real systems, such as cellular aqueous environments and soft matter materials, have in general high polymer concentrations. In these conditions, the intrachain coil-to-globule transition often occurs alongside interchain aggregation within a specific temperature range. This study aims to investigate the coupling between the intrachain coil-to-globule transition and interchain aggregation.

Experimental studies of the relationship between chain collapse and aggregation have been conducted under non-equilibrium conditions, specifically by quenching 
poly(methyl methacrylate) (PMMA) in solvents 
into the phase-separated regime~\cite{Maki2007,Nakata2007,Maki2008}. At low concentrations, individual chains first collapse, followed by aggregation of the collapsed chains into clusters of varying sizes. Within a specific concentration range, a sequential transition from chain collapse to chain aggregation is observed, with the extent of overlap between the two processes diminishing as concentration decreases. 
A prototypical example of polymers in aqueous solutions undergoing coil-to-globule phase transitions upon heating is poly($N$-isopropylacrylamide) (PNIPAM), which exhibits this transition around 305 K (32 °C)~\cite{Kubota1990,Graziano2000}. Depending on the concentration, PNIPAM can undergo either distinct or collective coil-to-globule and aggregation transitions~\cite{Ding2009}.

We study a model system of homopolymer aqueous solutions using molecular dynamics (MD) simulation and topological analysis. The model polymer solution is designed such that, at infinite dilution, a single polymer chain undergoes a gradual coil-globule transition around room temperature.~\cite{art:polymer_model} 
Here, we investigate the structural behavior of multiple homopolymer chains at finite concentration, focusing on the interplay between the coil-globule transition and their aggregation in water.

These conformational and collective transitions naturally raise important questions about the underlying topology of polymer chains and their interactions. Understanding the structural organization of multi-chain systems requires a framework capable of describing not only spatial configurations but also the connectivity and entanglement of chains.

The two main approaches to topologically describe a system of linear chains such as proteins, peptides, or RNA molecules, are knot theory~\cite{Adams2004,Lim2015} and circuit topology (CT)~\cite{Mashaghi2014,Mugler2014,Scalvini2020}. Both frameworks decompose a set of entangled polymers into fundamental units. For CT, these fundamental units are topological ``motifs'' formed by the {\color{black} mutual relation between two contact pairs}, while knot theory decomposes the system into {\color{black}combinations} of prime knots. Circuit topology serves as a complimentary approach to knot theory, since it is able to describe hard contacts (i.e., chemical or physical bonds between atoms, residues, etc.) for open chains, while knot theory focuses on entanglement and requires the chains to be closed. While CT can be extended to describe single-chain entanglement (soft contacts)~\cite{GOLOVNEV2020,Berx2023}, we will not focus on that aspect here. Since our system consists of open chains, we will therefore analyse it within the CT framework. Since CT mainly concerns the structures formed by interacting chains, it is ideally suited to study structural phase transitions~\cite{Berx2024}.


Let us set the stage for the polymer model studied in this work. We simulate $m$ linear freely-linked chains of $n$ spherical hydrophobic monomers with fixed intermonomer distances of $b = 0.345$ nm in the TIP4P/2005 water model~\cite{art:TIP4P2005}. The simulated systems comprise {\color{black}either $m=4$ or $m=8$} polymer chains with $n=30$ monomers each, together with 8000 water molecules. {\color{black} For $m=4$, the representative number density of monomers (at 298 K and 1 bar) is 0.49 nm$^{-3}$ 
and the corresponding concentration is 0.81~mol/L, and for $m=8$ they are doubled. }

The potential energy of the system is the sum of monomer-monomer, 
monomer-water, and water-water pair potentials. The first two pair potentials 
are Lennard-Jones (LJ) potentials, 
\begin{equation}
    \label{eq:LJ}
    \phi_{\rm{LJ}}(r) = 4\epsilon \left[ \left( \frac{\sigma}{r} \right)^{12} - \left( \frac{\sigma}{r} \right)^{6} \right]\,,
    \end{equation}
while the water-water pair potential is a sum of the LJ potential for the pair 
of oxygen sites and Coulomb potentials for the charged sites, as described by the TIP4P/2005 model.
The cut-off distance for the LJ potential was set to 1.1 nm, while the Coulomb potential was evaluated by the particle mesh Ewald method with the real space cut-off 
distance of 1.1 nm. 

The LJ parameters, $\epsilon_{\rm m}$ and $\sigma_{\rm m}$, 
for the monomer-monomer interactions are those for methane in the TraPPE-UA model~\cite{art:TraPPE-UA}:
$\epsilon_{\rm m}=1.230$~kJ~mol$^{-1}$ and $\sigma_{\rm m}=0.3730$~nm. 
The LJ parameters $\epsilon_{\rm wm}$ and $\sigma_{\rm wm}$ for the oxygen (of H$_2$O) -monomer pair interaction are $\epsilon_{\rm{wm}}$=1.356 kJ mol$^{-1}$ and $\sigma_{\rm{wm}}$=0.3444 nm. With these LJ parameters the simple model polymer chain 
in the TIP4P/2005 water undergoes the coil-globule transition near room temperature~\cite{art:polymer_model}. 
{\color{black} We confirmed that for both monomer concentrations of 0.81 and 1.62 mol/L, water and individual monomers mix at all temperatures from 240~K to 360~K.}

MD simulations were performed using GROMACS 2018~\cite{art:GROMACS} in the isothermal-isobaric ($NpT$) ensemble, using periodic boundary conditions and a time step of 1~fs. The coordinates were sampled every 50~fs. The pressure was maintained at 1~bar using the Parrinello-Rahman method, and the temperature was controlled using the Nos\'e-Hoover method. The simulation time was 50~ns after equilibration at each temperature.

\begin{figure}
    \centering
    \includegraphics[width=\linewidth]{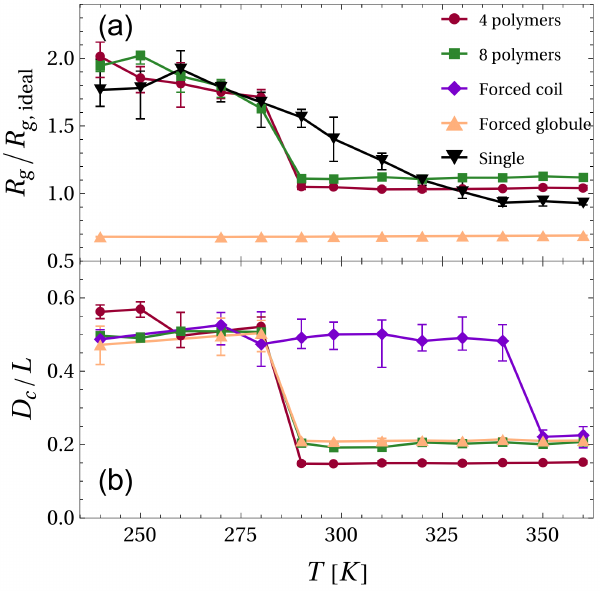}
    \caption{The radius of gyration $R_{\rm g}$ {\bf (a)} and degree of cohesion $D_{\rm c}$ {\bf (b)} as a function of temperature for the simulated models with {\color{black} $4$ (red) and $ 8$ (green) }polymers. For reference, the single-chain values are also shown (black). $R_{\rm g}$ and $D_{\rm c}$ are rescaled by the radius of gyration of the ideal chain $R_{\rm g, ideal} = b\sqrt{(n-1)/6}$ and system size $L$, respectively. }
    \label{fig:Rg_Dc}
\end{figure}

Together with topological measures we use two geometric ones: the radius of gyration, $R_{\rm g}$, a measure of the compactness of a polymer chain, and the degree of aggregation, $D_{\rm c}$, a measure of the inhomogeneity of the polymer solution. For a polymer with $n$ monomers at coordinates $\mathbf{r}_i$, 
\begin{equation}
    \label{eq:radius_gyration}
    R_g = \sqrt{\frac{1}{n} \left< \sum_{i=1}^{n} (\mathbf{r}_i - \mathbf{r}_{\rm c})^2\right>},
\end{equation}
where $\langle \cdots \rangle$ denotes the ensemble average over all polymers and the time average. The progress of the coil-globule transition is measured by the temperature dependence of $R_{\rm{g}}$. Similarly, the degree of aggregation of polymers is evaluated by 
\begin{equation}
    \label{eq:dc}
    D_{\rm c}= 
    \left< |\mathbf{r}^\alpha_{\rm c}-\mathbf{r}^\beta_{\rm c}|\right>,
\end{equation}
where $\alpha$ and $\beta$ are indices assigned to the polymers, 
and $\langle \cdots \rangle$ now represents the average over all polymer pairs in water and over time. We note that $D_{\rm c}$ for an ideal-gas model is proportional to $L$, the side length of the simulation box under periodic boundary conditions. Therefore, we will present the dimensionless quantity $D_{\rm c}/L$ in Figure \ref{fig:Rg_Dc}(b).

For comparison purposes, in addition to the model polymer solution, we examine the forced-coil systems and the forced-globule system. For the former, we fix the polymer configuration by imposing constraints on monomer distances and angles. Specifically, 
the inter-monomer distance is set to 2.949~nm for monomers numbered 1--11 and 20--30, and to 2.660~nm for monomers numbered 6--15, 11--20, and 16--25. The angle formed by three points is set to 176.14~degrees for the triplets 1--11--20 and 11--20--30. 
Note that the numbering runs from one end of the chain to the other. 
For the forced-globule system, we increase the Lennard-Jones (LJ) energy parameter for monomer-monomer intrachain interactions from $\epsilon_{\rm{m}}$ to $10\,\epsilon_{\rm{m}}$, while keeping the parameter for interchain interactions unchanged, collapsing the polymer into a globular state.

\begin{figure}
    \centering
    \includegraphics[width=\linewidth]{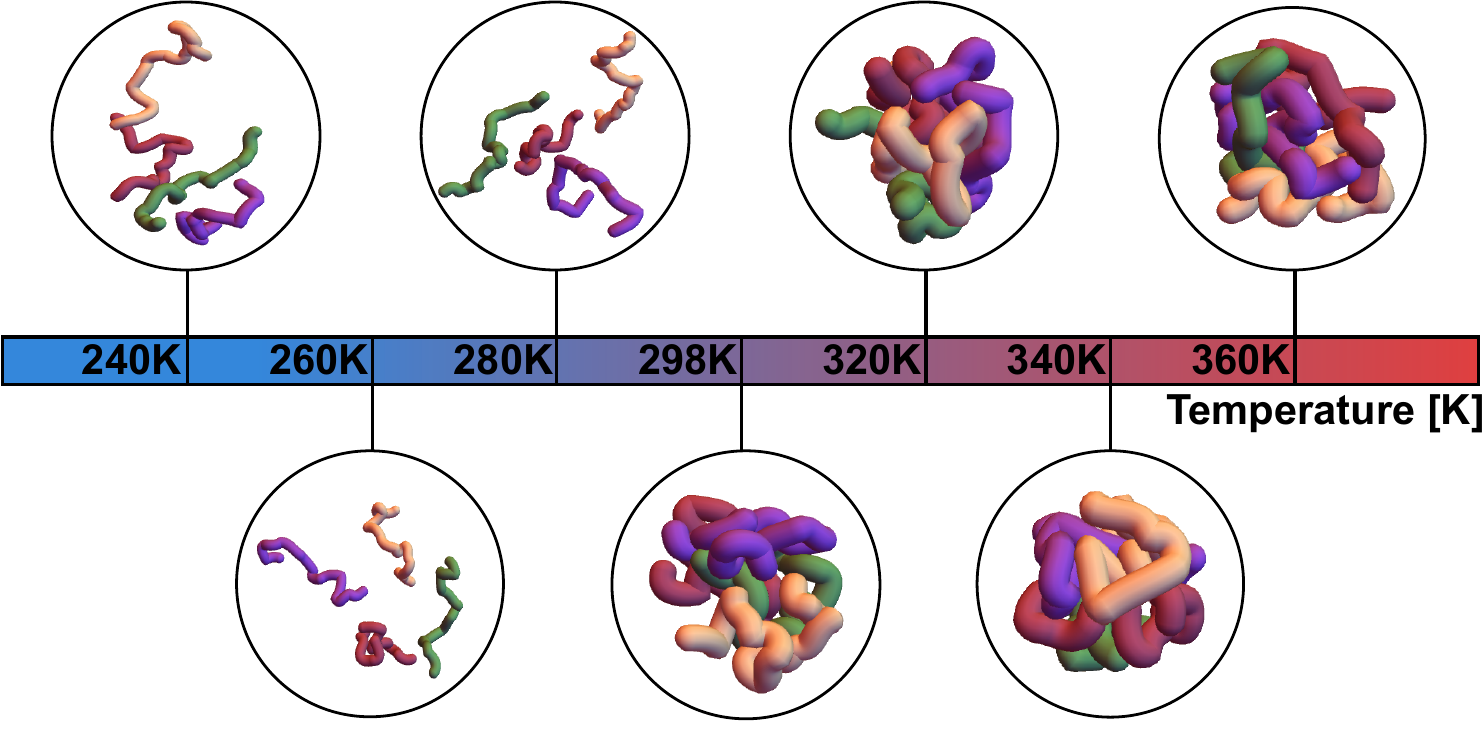}
    \caption{{\color{black}System configurations for different temperatures, showing the structural transition at a characteristic temperature $T_c \approx 290$, for $m=4$ polymers with $n=30$ monomers each. The polymers collapse from a dilute coiled state into a globular aggregated state.}}
    \label{fig:structures}
\end{figure}



For the circuit topological description of the system, let us first consider the arrangement of contacts on a single linear chain. This arrangement is a topological property, invariant to chain folding or stretching. In circuit topology (CT), motifs are defined as pairwise topological arrangements of contacts, e.g., contacts $\alpha$ and $\beta$.

Contacts $\alpha$ and $\beta$ may involve $\mathcal{M} = 1$ to $4$ chains. CT provides a second-order topological description of the system. At the first-order description, considering
individual contacts only, CT motifs are undefined, and contacts are classified as intrachain (within one chain) or interchain (between two chains). 
This analysis extends to tertiary, quaternary, or higher-order arrangements. As the number of distinct topological motifs grows exponentially with the number of contacts, we limit our analysis to binary arrangements of contacts, consistent with current practice.

\begin{figure}[htp]
\centering
\includegraphics[width=\linewidth]{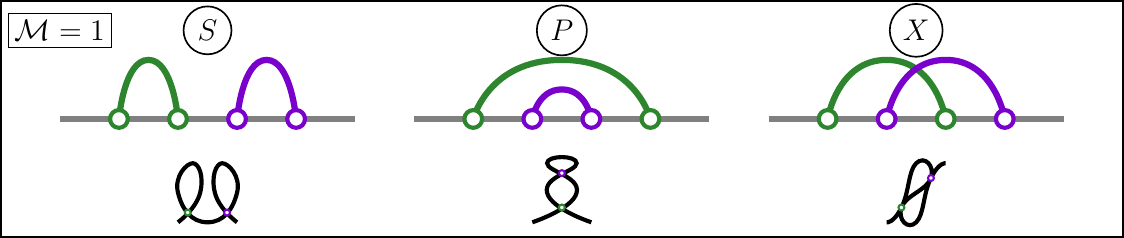} 
\caption{Table of single-chain circuit topology motifs. Contacts $\alpha$, $\beta$ are represented by green and purple lines connecting contact sites (white circles). Underneath each motif, a cartoon representation is shown illustrating the concomitant topology.}
\label{fig:CTtable_single}
\end{figure}

Each contact involves two monomer sites, and each motif, a topological arrangement of two contacts $\alpha$ and $\beta$, involves four monomers. For $\mathcal{M} = 1$, three motifs are possible: series (S), parallel (P), and cross (X). These are visualized in Fig.~\ref{fig:CTtable_single}. The motifs describe the topological relation between two loops in a single chain, classified as topologically independent (S) or bound (P and X).

\begin{figure}[htp]
    \centering
    \includegraphics[width=\linewidth]{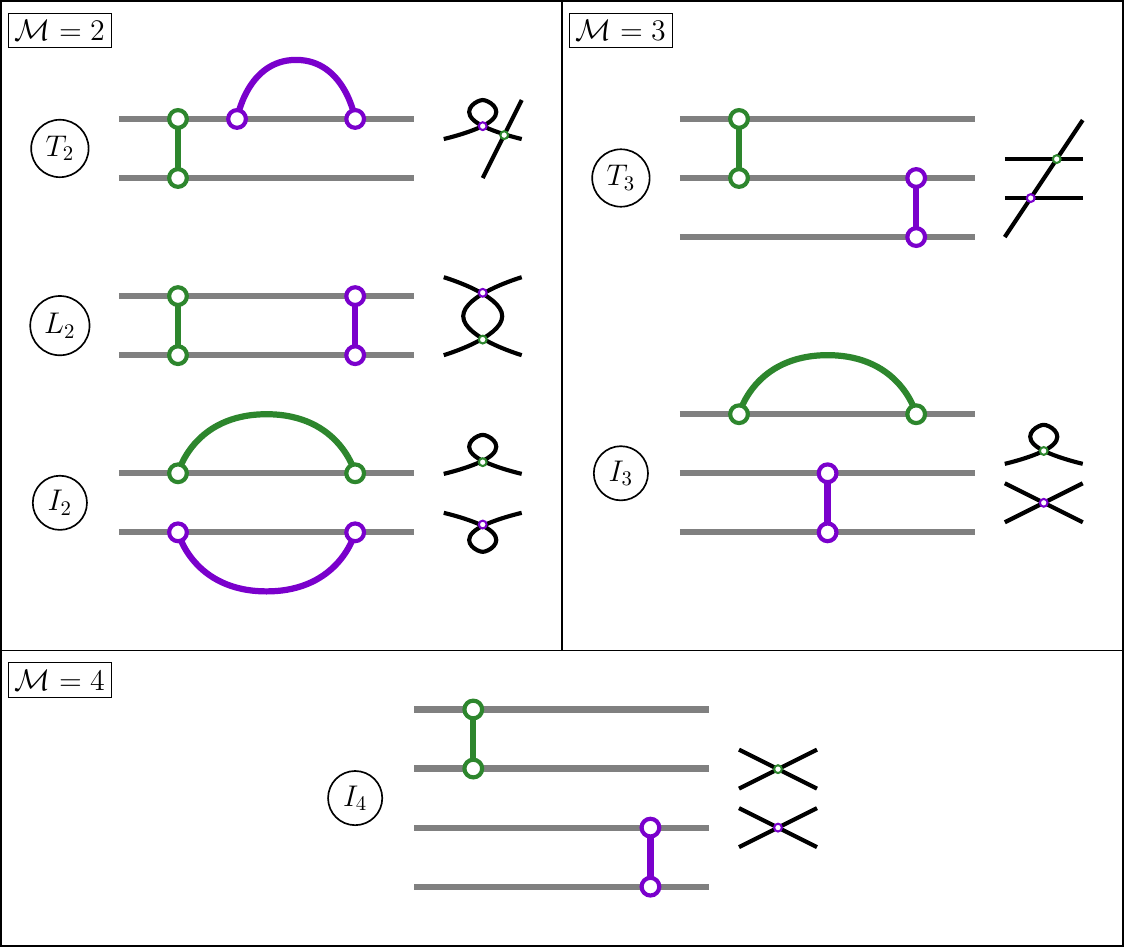} 
    \caption{Table of single-chain circuit topology motifs. Contacts $\alpha$, $\beta$ are represented by green and purple lines connecting contact sites (white circles). Next to each motif, a cartoon representation is shown illustrating the concomitant topology. Note that when a motif is degenerate only one example is shown.}
    \label{fig:CTtable_multi}
\end{figure}

In a multichain polymer system, additional motifs are identified by allowing contacts between distinct chains~\cite{Heidari2022,Berx2024}. Since four sites of two contacts can be distributed among different chains, the maximum number of chains in a single CT motif is $\mathcal{M}_{\rm max} = 4$. All multichain CT motifs are illustrated in Fig.~\ref{fig:CTtable_multi}, together with a simplified cartoon representation.

For $\mathcal{M}=2$, three motifs are identified. The tandem (T$_2$) motif is formed by one intrachain and one interchain contact, the loop (L$_2$) motif by two interchain contacts, and the independent (I$_2$) motif by two intrachain contacts, each on their respective polymer. Note that  I$_2$ is only determined topologically. The two chains can be geometrically distant or close,  provided each has an intrachain contact. 

For $\mathcal{M} = 3$, two motifs are identified: the tandem (T$_3$) motif, comprising two interchain contacts involving one common chain and two others, and the independent (I$_3$) motif, comprising one intrachain contact on one chain and one interchain contact between two others.  Finally, for $\mathcal{M} = 4$, only one motif exists: the independent (I$_4$) motif, with two interchain contacts, each between a distinct pair of chains. 

The system’s first-order description, i.e., the number of intrachain and interchain contacts, follows from the exact counts of CT motifs. Let $N_{\rm intra}$ and $N_{\rm inter}$ denote the number of intrachain and interchain contacts, respectively, in the system. The total number of contacts is $N = N_{\rm intra} + N_{\rm inter}$. The CT motif counts are denoted $S$, $P$, $X$, $T_2$, $L_2$, $I_2$, $T_3$, $I_3$, $I_4$. Their fractions are denoted by corresponding lowercases: $s$, $p$, $x$, $t_2$, $l_2$, $i_2$, $t_3$, $i_3$, $i_4$. 

When there are $N$ contacts in a multichain polymer system, a given intrachain contact generates $N-1$ motifs involving at least one intrachain contact, i.e., S, P, X, I$_2$, T$_2$ and I$_3$. 
As there are $N_{\rm intra}$ intrachain contacts,  $(N-1)N_{\rm intra}$ would be the total number of such motifs. But the motifs S, P, X, I$_2$ involve two intrachain contacts and therefore they are doubly counted in $(N-1)N_{\rm intra}$. 
Consequently, $(N-1)N_{\rm intra} =2(S+P+X+I_2) + T_2 + I_3$, or, equivalently,
\begin{equation}
    \label{eq:Nintra}
    N_{\rm intra} = \frac{1}{N-1}\left[2(S+P+X+I_2) + T_2 + I_3\right]\,.
\end{equation}
The fraction of intrachain contacts $r_{\rm intra} = N_{\rm intra}/N$ is then 
\begin{equation}
    \label{eq:nintra}
    \begin{split}
    r_{\rm intra} &= \frac{1}{N(N-1)}\left[2(S+P+X+I_2) + T_2 + I_3\right] \\
    &= \frac{1}{M}\left(S+P+X+I_2\right) + \frac{1}{2M}\left(T_2 + I_3\right)\\
    &= (s+p+x+i_2) + \frac{1}{2}(t_2 + i_3)\,,
    \end{split}
\end{equation}
where in the second line we have used the fact that the total number of CT motifs is $M = N(N-1)/2$. Similarly, the fraction of interchain contacts is given by
\begin{equation}
    \label{eq:ninter}
    r_{\rm inter} = (l_2 + t_3 + i_4) + \frac{1}{2}(t_2 + i_3)\,.
\end{equation}

\begin{figure}[htp]
    \centering
    \begin{subfigure}{0.49\linewidth}
        \includegraphics[width=1\linewidth]{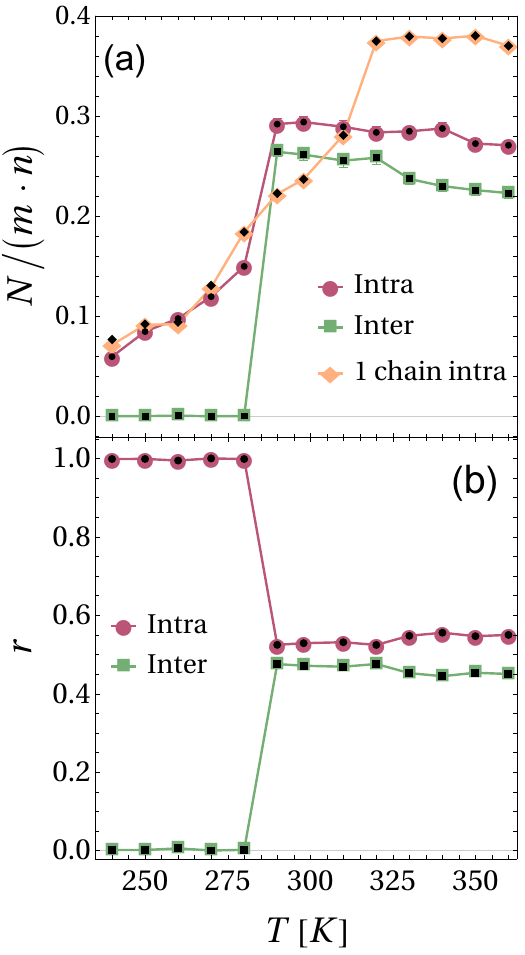} 
    \end{subfigure}
    \begin{subfigure}{0.49\linewidth}
        \includegraphics[width=1\linewidth]{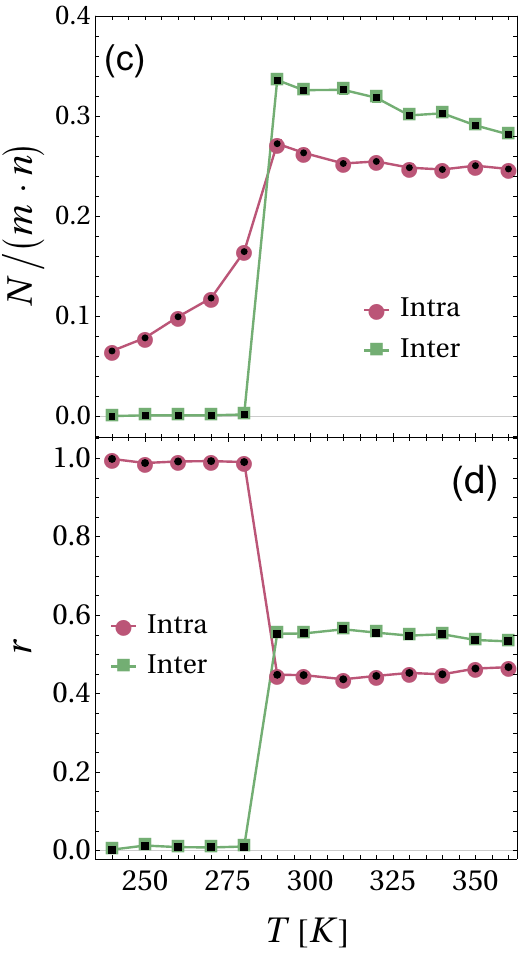}
    \end{subfigure}
    \caption{\color{black}Average numbers of intrachain contacts $N_{\rm intra}$ (red circles) and interchain contacts $N_{\rm inter}$ (green squares), normalized by the total number of monomers $m\cdot n$, together with their respective fractions $r_{\rm intra}$ (red circles) and $r_{\rm inter}$ (green squares), as functions of temperature $T$.
{\bf(a)} $N_{\rm intra}$ and $N_{\rm inter}$ for the 4-polymer system, with $N_{\rm intra}$ from the single-chain system (orange diamonds) shown for comparison;
{\bf(b)} $r_{\rm intra}$ and $r_{\rm inter}$ for the 4-polymer system;
{\bf(c)} $N_{\rm intra}$ and $N_{\rm inter}$ for the 8-polymer system;
{\bf(d)} $r_{\rm intra}$ and $r_{\rm inter}$ for the 8-polymer system.
Black (inner) symbols denote values obtained from eqs.~\eqref{eq:nintra} and \eqref{eq:ninter}, confirming the direct numerical calculations. For $T < T_c$, no interchain contacts are present and $N_{\rm intra}$ grows steadily, while for $T \geq T_c$, all measures ($N_{\rm intra}$, $N_{\rm inter}$, $r_{\rm intra}$, $r_{\rm inter}$) converge to constant values in both systems.}
    \label{fig:Ncontacts}
\end{figure}

It is thus straightforward to see that CT provides a more detailed level of description than simply classifying contacts as either intra- or interchain, but that the latter can be recovered through simple combinatorics. 
The number and relative fraction of contacts of both types changes drastically as a function of temperature. In Fig.~\ref{fig:Ncontacts}, we show the exact number of contacts and contact fractions as a function of temperature $T$, and compare the results with results obtained through the enumeration of CT motifs according to Eqs.~\eqref{eq:nintra} and~\eqref{eq:ninter} (black symbols). The results coincide exactly.

\begin{figure}[htp]
    \centering
    \includegraphics[width=\linewidth]{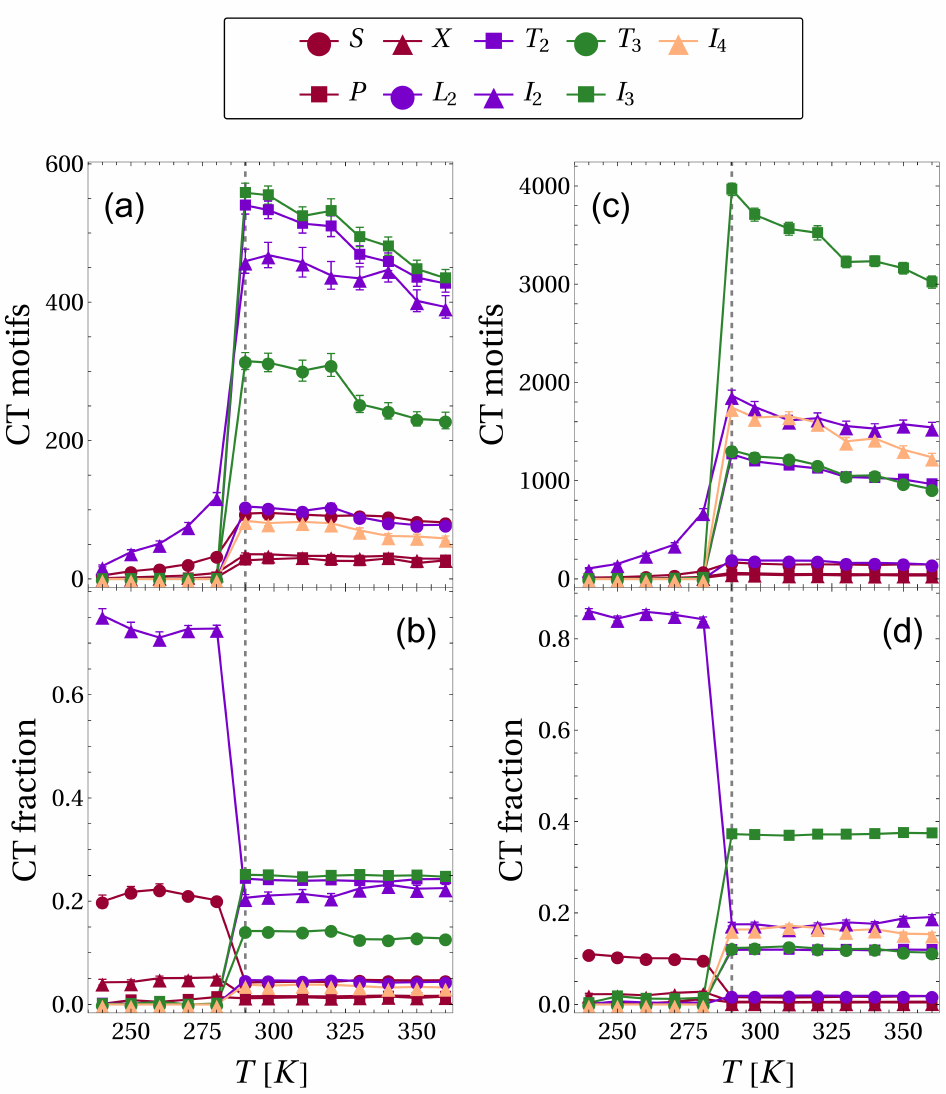}
    \caption{The average number and average fraction of CT motifs as a function of temperature. Single-, double- and triple-chain motifs are respectively coloured red, purple and green, while the quadruple-chain motif ($I_4$) is shown in orange. {\color{black}{\bf(a,c)} CT motifs for respectively the 4 and 8 polymer systems.} {\color{black}{\bf(b,d)} CT fractions for respectively the 4 and 8 polymer systems}. The structural phase transition at $T_c = 290$~K (dashed line) is easily seen in all motif numbers.}
    \label{fig:abs_CT}
\end{figure}

We now present our numerical results and examine single-polymer conformation, multichain configuration, and their relationship by tracking the radius of gyration ($R_{\rm g}$), interchain distance ($D_{\rm c}$), and CT motif fractions as a function of temperature (240 K to 360 K), as shown in Figs.~\ref{fig:Rg_Dc} and~\ref{fig:abs_CT}, respectively.

Figure~\ref{fig:Rg_Dc}(a) shows that, for both systems with $m=4$ and 8 polymer chains, $R_{\rm g}$ gradually decreases with increasing temperature $T$ from 240 to 280~K and drops {\color{black} sharply} at 290~K. Therefore, the structural change of each polymer at 290~K may be termed the coil-to-globule transition. 
Below this temperature, the polymers are extended and well‐solvated, which is
consistent with the fact that the osmotic second virial coefficient for hydrophobic molecules in water is {\it positive} at low temperatures.~\cite{koga2013jpcb} 
The system exhibits mostly the single-chain motifs S, P and X, and I$_2$, all of which increase their counts gradually as $T$ increases from 240 to 280~K, as shown in Fig.~\ref{fig:abs_CT}(a). The system is enriched in I$_2$ and S, indicating that loops are formed independently; if two loops are formed on the same chain they are far enough apart such that they do not interact. This is a distinct property of the coiled state of the polymers: they are stretched out spatially and form loops only locally, leading to a large radius of gyration. The gradual increase in single-chain motifs indicates that, as $T$ increases from 240 to 280~K, more loops form, which can interact by looping back onto themselves, sometimes forming the X motif. Since the contraction of a polymer is accompanied by loop formation, $R_{\rm g}$  decreases; it does so slowly because the conformational change of each polymer chain occurs independently in this temperature range. Figure~\ref{fig:Ncontacts}(a) also shows that $N_{\rm intra}$ for the multi-chain system gradually increases up to 280 K just as does $N_{\rm intra}$ for the single-chain system.

At a characteristic temperature  $T_{\rm c} \approx 290$~K, however, 
$N_{\rm intra}$ {\color{black} sharply} increases [Fig.~\ref{fig:Ncontacts}(a),(c)], $R_{\rm g}$ {\color{black} sharply} decreases (Fig.~\ref{fig:Rg_Dc}(a)), and the number of single-chain motifs increases drastically. These indicate that the polymers collectively undergo a coil-to-globule transition at $T_{\rm c}$ as they aggregate due to hydrophobic interactions. Conversely, $R_{\rm g}$ and $N_{\rm intra}$ for the single-chain system exhibit no discontinuous behavior over the whole range. In essence, inter-polymer effective interactions in water, i.e., the hydrophobic interactions, sharpen an otherwise gradual coil-to-globule conformational change. For comparison, the $R_{\rm g}$ values for two model systems containing either four forced-coiled or forced-globular polymers are also plotted in Figure~\ref{fig:Rg_Dc}(a).

Figure~\ref{fig:Rg_Dc}(b) shows how $D_{\rm c}/L$ varies with temperature for the four model systems. The $D_{\rm c}/L$ values for the standard and forced-globule systems exhibit a {\color{black} sharp} decrease as $T$ increases from 280 to 290~K, indicating that aggregation occurs sharply at approximately $T_{\rm c} \approx 290$~K. {\color{black} We also note that the aggregation temperatures for the two systems with $m=4$ and 8 coincide, indicating that $T_{\rm c}$ is only weakly dependent on the monomer concentration in the examined range. }
Above $T_{\rm c}$, the polymers have a reduced solvent-accessible surface area due to both the coil-to-globule transition and aggregation. 
This is readily observed in the CT spectrum (Figs.~\ref{fig:Ncontacts} and \ref{fig:abs_CT}): for $T < T_{\rm c}$, no motifs exhibiting interchain interactions can exist ($N_{\rm inter}=0$). At $T_{\rm c}$, $N_{\rm inter}$ sharply jumps to a non-zero value, and $N_{\rm intra}$ also rises significantly, suggesting a collective collapse-aggregation transition. With increasing temperature, the total number of contacts---and thus the CT motifs---decreases, but their relative fractions remain stationary. 
This indicates that the system reaches a stable macroscopic configuration where the topology remains similar, but thermal fluctuations can disrupt existing contacts, allowing individual chains to rearrange in confined environments. {\color{black}
Additionally, the hydrophobic effect, manifested as low solubility and hydrophobic monomer-monomer attraction, may be maximal around $T_{\rm c}$. Beyond this point, more water molecules may penetrate the polymer aggregate, thereby reducing intra- and inter-monomer contacts.}
For $T>T_{\rm c}$ the motifs I$_3$ and T$_2$ with a single interchain bond dominate, closely followed by I$_2$,
indicating that loops formed before the aggregation transition persist and are supplemented by interchain interactions. This reduces the relative fraction of all single-chain CT motifs, including I$_2$.

There are several notable differences between our results and previous studies on semiflexible polymers in which temperature is held constant while the Lennard-Jones interaction strength $\epsilon$ and polymer stiffness $\kappa$ are varied~\cite{Heidari2022,Berx2024}. In particular, those studies employed an implicit solvent model and did not include hydrophobic interactions. Within that framework, increasing $\epsilon$ effectively corresponds to decreasing temperature in our work. However, unlike our model, such an increase in interaction strength drives the coil-to-globule transition. Moreover, by increasing the stiffness and reducing the length of the individual polymers, such systems exhibit an isotropic-nematic transition wherein individual polymers align according to the system's global director, forming bundles or tactoids. This directed rearrangement of polymers is accompanied by an enrichment in T$_3$, L$_2$ and I$_4$ motifs, which form prior to possible chain collapse. In contrast, our system is dominated by the I$_3$, I$_2$ and T$_2$ motifs.

For the forced-coil system, however, the polymers aggregate at 350~K, a much higher temperature. Therefore, we conclude that the temperature of phase separation, driven by hydrophobic interactions, strongly depends on polymer 
conformation (290~K vs.\ 350~K). 

We note that for $T > T_{\rm c}$, $D_{\rm c}/L$ for the standard system is lower than that for the forced- globule system. This reflects the fact that forced-globular polymer chains cannot snake trough the aggregate as flexible standard polymer chains do. A CT analysis reveals that this is indeed the case; the number of single-chain motifs and $I_2$ remains approximately constant for the entire temperature range, indicating that loops are stable and that for $T > T_{\rm c}$ they are simply supplemented by interchain contacts, increasing only motifs of type I$_3$ and T$_2$.

Below $T_{\rm c}$, where the solution is homogeneous, 
$R_{\rm g}$ decreases by 15\% as $T$ increases from 240 to 280~K. 
Does this decrease in $R_{\rm g}$ upon heating promote phase separation? In other words, if hypothetical polymer chains are used whose $R_{\rm g}$ remains  the same as that for the standard system at 240 K, would the aggregation temperature be higher than $T_{\rm c}$? To answer this question directly, we would need to evaluate $T_{\rm c}$ for the hypothetical polymer chains in water. Without such simulations, however, we propose that the answer is likely affirmative. First, as confirmed earlier and shown in Figure~\ref{fig:Rg_Dc}(b), the aggregation temperature for the forced-coil  system, which has a larger $R_{\rm g}$ than the standard system, is 350~K, substantially higher than 290~K. This indicates that the aggregation temperature is highly sensitive to polymer conformation. Second, although $R_{\rm g}$ decreases by only 15\% from 240 to 280~K, the conformational change in the standard polymers over this temperature range is significant in terms of the number of intrachain contacts, $N_{\rm intra}$, and the corresponding CT motif, $I_2$, as shown in Figures~\ref{fig:Ncontacts} and~\ref{fig:abs_CT}(a): $N_{\rm intra}$ increases by 152\% and $I_2$ by 533\%.


In this work, we have investigated the structural and topological behavior of interacting hydrophobic polymer chains in aqueous solutions by combining molecular dynamics simulations with a CT analysis. By focusing on hard contacts in open chains, CT provided a robust framework for classifying and quantifying intra- and interchain interactions, complementing conventional geometric descriptors such as radius of gyration and degree of aggregation.

Our results reveal a temperature-driven collective structural transition, wherein homopolymer chains shift from isolated coil states to aggregated globule states. The temperature dependence of the radius of gyration ($R_{\rm g}$) in the multi-chain system shows that interchain hydrophobic interactions drive the complete collapse of polymer chains, coinciding with the onset of aggregation at the same temperature. Comparison of the standard system with two constrained systems, where interchain monomer-monomer interactions are identical, indicates that the aggregation temperature increases with greater chain extension. This suggests that the preliminary collapse of polymer chains (a 15\% reduction in $R_{\rm g}$ from 240 to 280 K) lowers the aggregation temperature compared to a hypothetical system where no reduction in $R_{\rm g}$ occurs. 

This transition is captured not only by the standard geometric observables, but also in the emergence and redistribution of CT motifs. We showed that CT motif enumeration not only recovers first-order descriptors (i.e., intra- and interchain contact fractions) through simple combinatorics, but also affords a higher-order understanding of polymer organization inaccessible to traditional geometric metrics alone. In particular, the transition from predominantly intrachain motifs (S, P, X) at low temperatures to a significant population of multichain motifs (T$_2$, L$_2$, T$_3$, I$_3$, I$_4$) at higher temperatures correlates directly with the emergence of aggregated states in the system. {\color{black} The sharp onset of interchain contacts and motifs should be interpreted cautiously. Observed in finite-size, explicit-solvent simulations, this behavior does not necessarily imply a breakdown of Ising-class criticality. It is possible that the polymer concentration is not sufficiently close to the critical concentration, or the temperature interval of 10 K is too large.
To determine whether or not the step-like behavior persists in the thermodynamic limit, systematic finite-size scaling and/or numerical simulations of simpler model systems---such as lattice models, implicit-solvent models, or explicit-but-simplified solvent models---are required.
}


{\color{black} 
The model polymer in this study is a freely jointed chain of hydrophobic monomers. Incorporating hydrophilic monomers or groups into the polymer chain enables the study of chemical inhomogeneity effects on collective phase transitions. 
The CT analysis-based approach, combined with explicit solvent-model MD simulations that are readily extensible to higher degrees of polymerization and non-linear chain architectures, facilitates investigation of co-solvent effects, such as those from alcohols and electrolytes, on phase transitions in aqueous polymer solutions. 
This approach opens new avenues for exploring hierarchical phase transitions in soft matter, particularly in biologically relevant systems like protein condensation, RNA aggregation, and cellular phase-separated compartments.
}

\begin{acknowledgments}
K.K. thanks JSPS KAKENHI (Grant Numbers 18KK0151, 20H02696, and 25K00969). Part of the computation was performed using Research Center for Computational Science, Okazaki, Japan (Project:23-IMS-C112, 24-IMS-C106, and 25-IMS-C106). 
J.B. thanks Okayama University for generous hospitality and the Research Foundation of DPhil Ragna Rask-Nielsen for funding the project.
\end{acknowledgments}

\section*{Data Availability Statement}
The data that support the findings of this study are available from the corresponding author upon reasonable request.

\bibliography{biblio}

\end{document}